\documentclass[aps,prl,twocolumn,groupedaddress,superscriptaddress]{revtex4}

\usepackage{graphicx}
\usepackage{dcolumn}
\usepackage{bm}
\usepackage{xcolor}

\begin{document}

\title{Quantum spin Hall effect on germanene nanoroad embedded in a completely hydrogenated germanene}

\author{L. Seixas}
\email[]{lseixas@if.usp.br}
\affiliation{Instituto de F\'isica,
Universidade de S\~ao Paulo, CP 66318, 05315-970, S\~ao Paulo, SP, Brazil.}

\author{J. E. Padilha}
\email[]{padilha@if.usp.br}
\affiliation{Instituto de F\'isica,
Universidade de S\~ao Paulo, CP 66318, 05315-970, S\~ao Paulo, SP, Brazil.}

\author{A. Fazzio}
\email[]{fazzio@if.usp.br}
\affiliation{Instituto de F\'isica, Universidade de S\~ao Paulo, CP 66318, 05315-970, S\~ao Paulo, SP, Brazil.}

\date{\today}


\begin{abstract}
 We show that germanene nanoroads embedded in a completely hydrogenated germanene (germanane) exhibits a Quantum Spin Hall Effect (QSHE). These nanoroads can be obtained experimentally by local hydrogen dissociation from germanane. Using first principle calculations we predict that germanene nanoroads with zigzag interfaces show dissipationless conducting channels with in-plane and out-of-plane spin textures.
\end{abstract}

\maketitle

The search for topological quantum phases of matter led condensed matter scientists to inspect, under a different point of view, materials known as Topological Insulators (TI). Those materials exhibit different properties and physical phenomena, like the quantum spin Hall effect (QSHE) \cite{PRL95-226801-2005}, axion electrodynamics \cite{NaturePhysics.6.284} and Majorana fermions \cite{PRL.100.096407}. The QSHE in two dimensional (2D) materials was proposed by Kane and Mele for graphene \cite{PRL95-226801-2005}, followed by an independent proposal by Bernevig, Hughes and Zhang for the CdTe/HgTe/CdTe quantum well \cite{science314-1757-2006}. The latter has been confirmed experimentally \cite{science318-766-2007}, but for graphene due to the small spin-orbit bandgap, the observation of the QSHE would apply only at unaffordable experimental conditions \cite{PRB74-165310-2006,PRB75-041401-2007}. 

New candidate materials has been proposed to overcome the issue presented by graphene for the observation of the QSHE. Among some of these materials are the silicene \cite{APL96-183102-2010, APL97-223109-2010, APE5-045802-2012, NanoLett12-3507-2012, PRL108-245501-2012, SSR67-1-2012}, germanene \cite{PRL107-076802-2011}, 2D hexagonal Si$_{x}$Ge$_{1-x}$ \cite{PRB.88.201106}, and stanene \cite{PRL.111.136804}. They present similar properties to graphene, having some additional features, such as: (i) buckled honeycomb lattice; (ii) relatively large spin-orbit coupling (SOC), with bandgap of the order of meV and topological invariant $\mathbf{Z}_2$ = 1 \cite{PRL107-076802-2011, PRB.88.201106}. Such characteristics makes these 2D TI  good candidates for the observation of the QSHE. 

From the experimental point of view, the pursuit of 2D materials that could display the QSHE was intensified by the synthesization of silicene. Silicene is mostly grown on metal surfaces,\cite{APL96-183102-2010, APL97-223109-2010, APE5-045802-2012, NanoLett12-3507-2012, PRL108-245501-2012, SS608-297-2013, NanoLett13-685-2013, PRL110-076801-2013, PRL110-085504-2013} that introduces some difficulties to isolate this single layer material and construct devices that will allow the measurement of the QSHE. Also the SOC bandgap in silicene is small, around 1.9 meV, so that the temperature to observe the QSHE will also be very low.  

A recent work of Bianco \textit{et al.} \cite{ACSNano7-4414-2013} reported synthesis of germanane, a completely hydrogenated germanene structure. This opens up a new route to the experimental observation of the QSHE at higher temperatures. The formation and stability of this material, as well as the study of the QSHE may play an important role in nanoelectronics \cite{ACSNano7-4414-2013}.

In this work we propose a feasible setup for the observation of QSHE that is based on germanane \cite{ACSNano7-4414-2013}. The system is a germanene nanoroad embedded in germanane. We find that  the germanene nanoroads with zigzag interfaces presents dissipationless conducting channels with in-plane and out-of-plane spin texture. We also show that with a small perturbation at the edge is possible to lift the degeneracy between interfaces, allowing a single-interface conduction. The electronic structure simulations were performed within the density functional theory.

As in graphane \cite{PRB75-153401-2007}, a completely hydrogenated graphene, we can built a germanene nanoroad by removing the hydrogen atoms \cite{ACSNano4-6146-2010, NanoLett9-1540-2009, JAP109-054314-2011, JAP110-063715-2011, Nanotechnology.24.495201} in a desired pattern by, for example, hydrogen dissociation via plasma etching. This germanene structure inside germanane is a non-trivial topological insulator embedded in a trivial insulator and from the bulk-edge correspondence, gapless topological edge states should appear on this line-interface region. We employ parameter-free calculation to obtain the formation energies and the electronic structure of zigzag and armchair germanene nanoroads embedded in germanane.

First-principle calculations were performed within the framework of density functional theory (DFT) \cite{PhysRev136-B864-1964, PhysRev140-A1133-1965} as implemented in the \textsc{OpenMX} code \cite{openmx} for electronic band structures, and \textsc{Vasp} code \cite{PhysRevB.54.11169, PhysRevB.59.1758} for spin texture of Bloch states. For the exchange-correlation functional we used the GGA-PBE approximation \cite{PRL77-3865-1996}. The spin-orbit interaction in the electronic structure calculations was included via norm-conserving fully relativistic $j$-dependent pseudopotentials scheme, in the non-collinear spin DFT formalism \cite{PRB64-073106-2001}. All systems were fully relaxed with a force criteria of  0.001 eV/\AA. The spin texture was calculated for the interface states by the spin polarization vector
\begin{equation}
  \textbf{P}_n(\textbf{k}) = \langle u_{n\textbf{k}}^{\sigma} | \boldsymbol\sigma | u_{n \textbf{k}}^{\sigma} \rangle,
\end{equation}
where $|u_{n \textbf{k}}^{\sigma} \rangle$ are the Bloch states, and $\boldsymbol\sigma  = \left( \sigma_x, \sigma_y, \sigma_z \right)$ is the Pauli matrices vector.

\begin{figure}[!ht]
  \includegraphics[width=8.5cm]{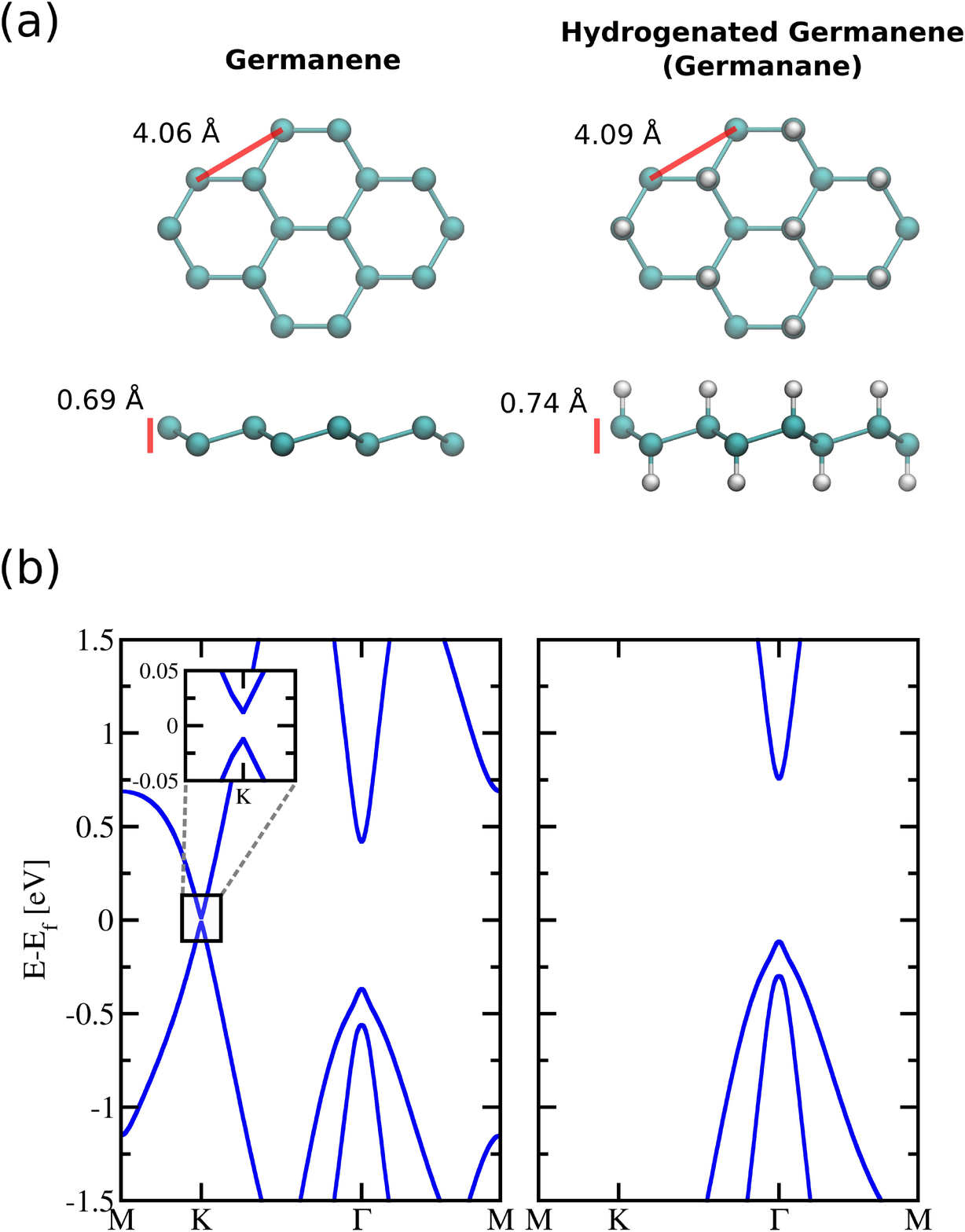}
  \caption{(a) Schematic representations of germanene (Left panel) and germanane (Right panel). (b) Electronic band structure of germanene (Left panel) and germanane (Right panel). The inset in the band structure of germanene shows the spin-orbit bandgap of 25 meV at $K$ point.}
  \label{Fig01}
\end{figure}

Figure \ref{Fig01}(a) shows the geometries for germanene and hydrogen decorated germanene (germanane). The buckling in germanane is $\sim 0.05$ \AA\, higher than in germanene. In Fig. \ref{Fig01}(b) we show the electronic band structure of germanene (Left panel) and germanane (Right panel). Germanene presents a spin-orbit band gap of 25 meV at $K$-point (see inset in Fig. \ref{Fig01}(b)) while the germanane is a semiconductor with direct band gap of 1.53 eV at $\Gamma$-point \cite{ACSNano7-4414-2013}. The $\mathbf{Z}_{2}$ index was evaluated using the Fu-Kane procedure \cite{PhysRevB.76.045302}, by counting the parity of the occupied bands at the time-reversed invariant momenta in the Brillouin zone. We obtained $\mathbf{Z}_{2} = 0$ for germanane (trivial insulator) and $\mathbf{Z}_{2} = 1$ for germanene (topological insulator).

In order to determine the formation energy and stability of the germanene nanoroads, we compare the energy to create a hydrogen vacancy ($V_{\mathrm{H}}$) and divacancy ($D_{\mathrm{H}}$) in germanane, with the formation energy of a germanium vacancy ($V_{\mathrm{Ge}}$) in germanene. Using single defects in a $4\times4$ supercell of germanane, we found that the formation energies of hydrogen vacancy and divacancy are $\Delta H_f(V_{\mathrm{H}})$ = 1.03 eV and $\Delta H_f(D_{\mathrm{H}})$ = 0.57 eV. For the formation energy of a germanium vacancy we obtain $\Delta H_f(V_{\mathrm{Ge}})$ = 2.36 eV. These results are comparable with the formation energies of hydrogen vacancy ($V_{\mathrm{H}}$) and divacancy ($D_{\mathrm{H}}$) in graphane, and carbon vacancy ($V_{\mathrm{C}}$) in graphene (for the carbon system the formation energies are: $\Delta H_f(V_{\mathrm{C}})$ = 8.04 eV, $\Delta H_f(V_{\mathrm{H}})$ = 1.98 eV, and $\Delta H_f(D_{\mathrm{H}})$ = 0.75 eV). From these results we can suggest that from an experiment with a controllable beam energy it is feasible to produce the desired pattern, without damaging the dehydrogenated germanene. Thus with a very controlled process, the germanene nanoroads would be stable in hydrogen dissociation processes, such as plasma etching.

\begin{figure}[!ht]
  \includegraphics[width=8.5cm]{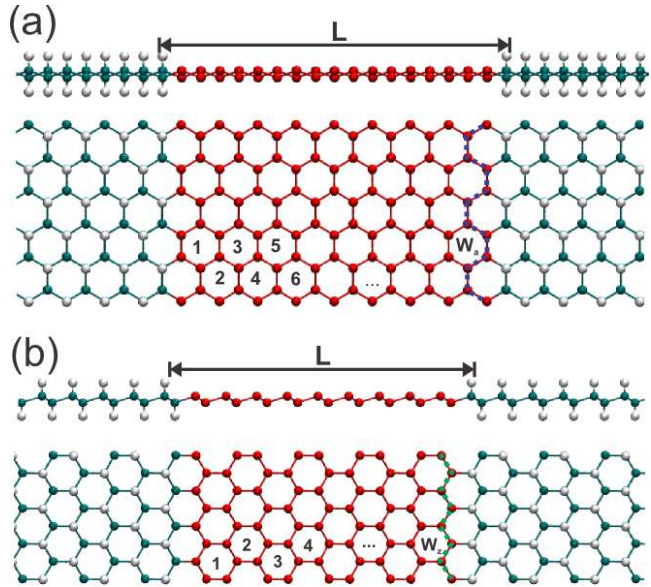}
  \caption{Schematic representations for: (a) armchair germanene nanoroad, (b) zigzag germanene nanoroad. Each representation are shown from sideview (Top panel) and topview (Bottom panel).}
  \label{Fig02}
\end{figure}

Based on the idea of local hydrogen dissociation, were assembled germanene nanoroads with armchair and zigzag interfaces, as shown in Fig. \ref{Fig02} (a) and (b). The widths of the germanene nanoroads are labeled by $W_a$ for armchair, and $W_z$ for zigzag nanoroads, where $W$ indicates the number of exposed hexagons of germanene in the unit cell, as shown in Fig. \ref{Fig02}. The values of these widths, $L$,  in \AA\, are given by
\begin{equation}
  L(W_z) = 1.14 + 3.56\, W_z,
\end{equation}
for zigzag nanoroads, and
\begin{equation}
  L(W_a) = 1.93 + 2.05\,  W_a,
\end{equation}
for the armchair nanoroads. Those nanoroads were studied using widths ranging from $W_a = 1$ up to $W_a = 32$ for the armchair, and from $W_z=1$ up to $W_z=17$ for the zigzag ones. 

\begin{figure}[!ht]
  \includegraphics[width=8.5cm]{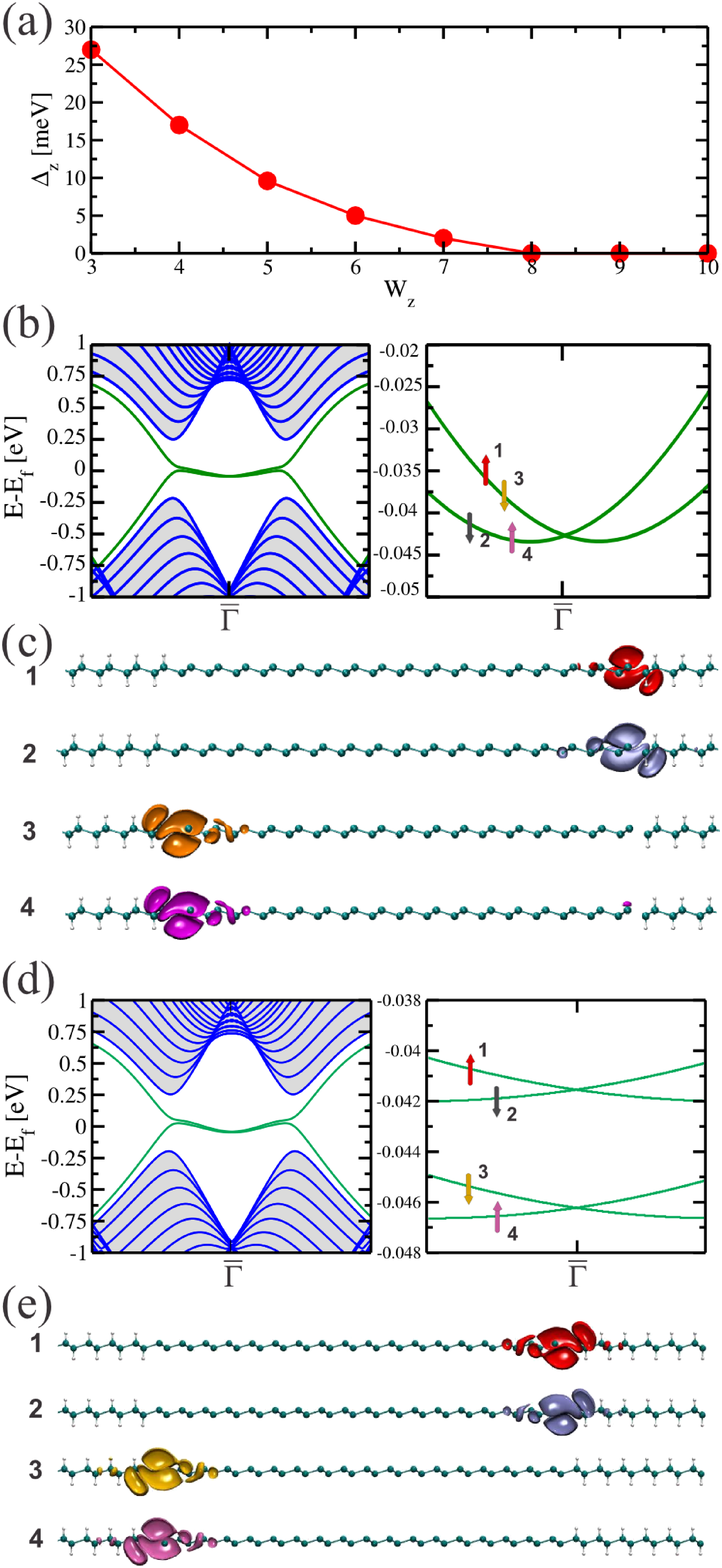}
  \caption{(a) Bandgap at $\overline{\Gamma}$ as function of width for zigzag nanoroad. (b) Right panel: Band structure of zigzag nanoroad for $W_z = 16$. Left panel: Edge states (green) with spin polarization for each band, labeled from 1 to 4. (c) Wavefunction isosurfaces near the $\overline{\Gamma}$ point for bands 1, 2, 3 and 4 in (b). (d) Right panel: Band structure of zigzag nanoroad, with width $W_z = 17$ and a small perturbation on the right edge. Left panel: Edge states (green) with spin polarization for each band, labeled from 1 to 4. (e) Wavefunction isosurfaces near $\overline{\Gamma}$ point for bands 1, 2, 3 and 4 in (d).}
  \label{Fig03}
\end{figure}

Due to the sublattice staggered potential for very narrow nanoroads, the interactions between interfaces could give rise to topologically assisted one-dimensional (1D) transport channels \cite{NanoLett2012-12.2936, PhysRevB.89.085429}. However, both germanene/germanane interfaces have the same valley-based topological indexes and also large SOC, splitting the bandgap up to $27$ meV for $W_z = 3$, as shown in Fig. \ref{Fig03}(a). For $W_{z} \leq 7$ all systems present a small bandgap and no zero energy topological interface state could be observed [See Fig. \ref{Fig03}(a)]. As the interaction between zigzag interface states decays exponentially \cite{PRB.88.121401}, we found that for $W_{z}>7$, the inter-edge interaction is no longer significant, leading to an one-dimensional metal states.

In Fig. \ref{Fig03}(b) we show the calculated electronic band structure for $W_z = 16$ on Left panel and on the Right panel we show a magnification of the edge states around $\overline{\Gamma}$ showing the spin polarization of each band. Each band is doubly degenerated, coming from each interface of the system. In Fig. \ref{Fig03}(c) we present the wave function with the corresponding spin polarization.  We can see that the first set of bands, 1 and 3, presents opposite spin polarizations and are located on different edges. The same happens with the second set of bands, 2 and 4. Such behavior is the main characteristic of the QSHE.
 
In Fig. \ref{Fig03}(d) we show the calculated electronic band structure for $W_z = 17$, where in this system we included a small perturbation in one of the edges. This perturbation was included to break the inversion symmetry between interfaces \cite{perturbation}. Now we can clearly observe two sets of interface states, that forms two one-dimensional and non-degenerate Dirac cones at the boundary of the Brillouin zone. We can also see the spin-locked polarization for each band (Right panel in Fig. \ref{Fig03}(d)) with the respective real space polarized wave function in Fig. \ref{Fig03}(e). 

\begin{figure}[!ht]
  \includegraphics[width=8.5cm]{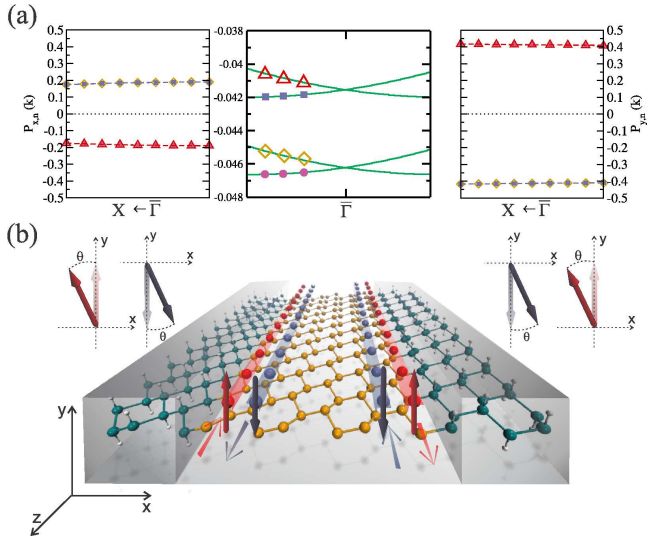}
  \caption{(a) Right panel: Spin texture component in the $x$ direction for bands: 1, 2, 3 and 4. Center panel: Symbolic convention for each band and $k$-points used in spin texture graphs. Left panel: Spin texture component in the $y$ direction for bands: 1, 2, 3 and 4. (b) Pictorial representation of spin texture of edge states. From spin polarization in the $x$ and $y$ directions, was inferred the polar angle $\theta = 25\,^{\circ}$ for spin polarization vector in spherical coordinate system.}
  \label{Fig04}
\end{figure}

The spin textures were calculated using the expectation values of Pauli matrices for Bloch states in the vicinity of the $\overline{\Gamma}$ point. The spin texture components $P_{x,n}$ and $P_{y,n}$ displays almost constant behavior, as shown in Fig. \ref{Fig04}. The interface states 1 and 4 (as labeled in Fig. \ref{Fig03}(b)) present $P_{x,n} \approx -0.2$ and $P_{y,n} \approx 0.4$, and the interface states 2 and 3 present $P_{x,n} \approx 0.2$ and $P_{y,n} \approx -0.4$. In the $z$-direction we found $P_{z,n} \approx 0$ for all interface states. Based on these component values, we can reconstruct the spin texture of the topological states, represented in Fig. \ref{Fig04}(b). For each interface state, the resulting spin polarization show slopes up to $\theta = 25\,^{\circ}$ with respect to the polar axis orthogonal to the material. We can infer that manipulating the electron chemical potential at meV scale or by a small interface-dependent perturbation that does not break the time reversal symmetry,  we could obtain electrical current passing through the interfaces, or a spin current through a single interface. The last one scenario could be seen as a leading signature for the germanene nanoroads, besides the in-plane component spin texture. This feature could be experimentally measured in a quantum Hall device in absence of any magnetic fields.

\begin{figure}[!ht]
  \includegraphics[width=8.5cm]{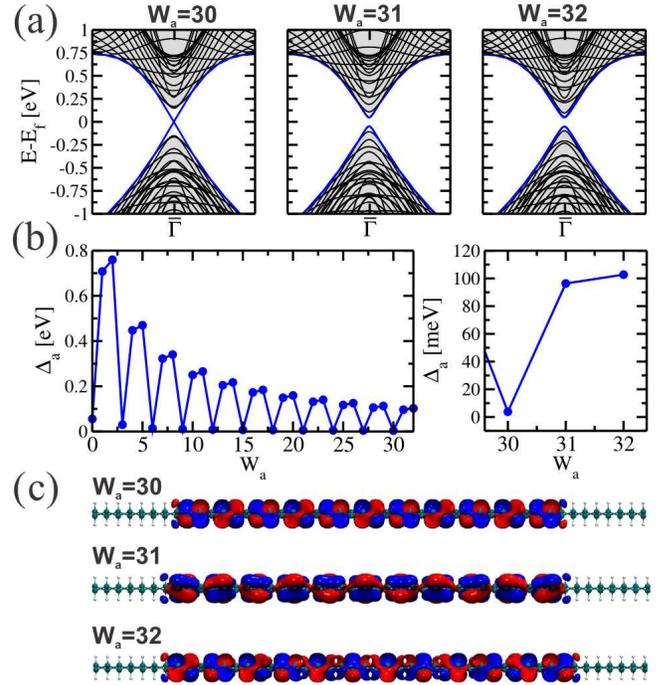}
  \caption{(a) Band structure of armchair germanene nanoroad with sizes from $W_a=30$ to $W_a=32$. (b) Right panel: Bandgap at $\overline{\Gamma}$ point as a function of width. Left panel: Bandgap for $W_a = 30, 31, 32$, showing three classes of electronic properties of armchair germanene nanoroads. (c) Wavefunction isosurfaces (HOMO) for positive (blue) and negative (red) values at $\overline{\Gamma}$ point for $W_a = 30, 31, 32.$}
  \label{Fig05}
\end{figure}

Finally, for the germanene nanoroads with armchair interfaces, we calculate the electronic band structure as function of width, as shown in Fig. \ref{Fig05}. The band structures present similarities in three distinct classes. In Fig. \ref{Fig05}(a) we show three representative systems, $W_a = 30$, $31$ and $32$. These classes of band structures show a bandgap fluctuation, with quasi-periodicity of $\Delta W_a = 3$, as shown in Fig. \ref{Fig05}(b). Despite the small bandgaps in the class $W_a = 3N$, where $N=1,2,\ldots$, all calculations show a non-zero energy gap within our width range. In Fig. \ref{Fig05}(c), we show that the wavefunctions (for the HOMO states) are delocalized throughout the nanoroad. This delocalization creates a long-range interaction that mix the two topological states. From recent results by Ezawa and Nagaosa \cite{PRB.88.121401} for silicene nanoribbons, the armchair edge is characterized by the absence of localized edge states, differently from zigzag that is characterized by localized edge states. In other words, the decay length of edge orbitals in armchair and zigzag termination are completely different.

In conclusion, we showed that the quantum spin Hall effect can be observed in germanene single nanoroads embedded in germanane at realistic experimental conditions. These nanoroads can be obtained by local hydrogen dissociation from germanane in the form of a strip (1D) with either zigzag or armchair interfaces. Beyond the QSHE at accessible temperatures, we found in-plane and out-of-plane spin polarization components which leads spin texture with polar slopes up to $25\,^{\circ}$, and non-degenerate interface states.

We would like to thank the Brazilian funding agencies: INCT/CNPq and FAPESP. We also thanks A. Janotti for a critical reading of an earlier version of this manuscript.


\begin{thebibliography}{99}
\bibitem{perturbation}
To create a small perturbation on the system with the aim to break the inversion symmetry between the interfaces we fixed, during the relaxation, the position of two hydrogen atoms neighbor to only one of the interfaces, leaving all the other atoms free to relax. The energy difference between the system fully relaxed and the one with the small perturbation is  $\sim 2$ meV.
\end{thebibliography}
\end{document}